# A Geometrical Modeller for HEP

R. Brun, A. Gheata
*CERN, CH 1211, Geneva 23, Switzerland*

M. Gheata
*ISS, RO 76900, Bucharest MG23, Romania*

For ALICE off-line collaboration

Geometrical modelling generally provides the geometrical description of a special structure and a set of services to "navigate" through it. HEP geometrical modellers are designed to handle high complexity detector geometries and they are usually embedded within simulation MC frameworks. The fact that these frameworks greatly depend on their specific geometrical tools makes simulation applications hardly portable to MC's other than the one they were designed for. The ALICE Off-line Project in collaboration with the ROOT team is proposing a multi-purpose geometrical modeller for HEP that is integrated within a virtual MC infrastructure. This tool has been optimised for performance with the geometry setups of several HEP experiments and provides a single representation for the geometry used by different applications such as simulation, reconstruction or event display.

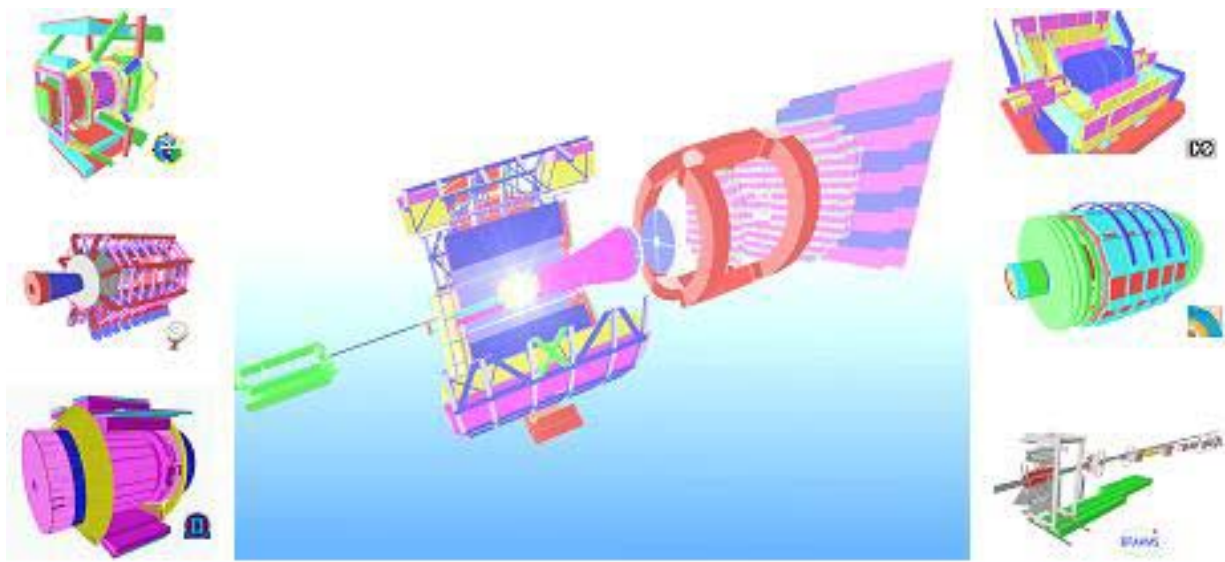

Figure 1: The ROOT geometry modeller is able to represent most HEP experiments

## 1. INTRODUCTION

The simulation of an experiment is one of the most important activities during its preparation as it can provide a better understanding of the detector response and how the acquired data will eventually look like. As HEP moves to new experiments with more and more complex detector systems, this becomes one of the most CPU time and resources consuming off-line task. It is handled by large simulation engines such as GEANT [5, 7] and has to share its geometry description with several other frameworks, such as reconstruction or event display.

Having one single geometry description may become a critical issue when several applications have to share related data and rely on detector module inter-dependent algorithms. In such cases having for instance a simplified geometry for reconstruction may not be an option anymore. We have also to consider the fact that geometry itself is a subject to changes in time. If several different implementations are used, it can be very difficult to maintain a consistent geometry. On the other hand the geometry description and navigation features are currently embedded and fully specific to the different simulation engines, therefore the general tendency is that several applications cluster in large frameworks around these in order to solve this problem. Since geometry modelling and related features are quite complex, simulation packages grow larger as new geometrical functionality is needed, the framework becoming less and less modular leaving no alternative for usage of other simulation packages other than the initial choice.

The geometry design itself is coded in a manner optimizing the simulation performance for the specific geometry modeller used, therefore becoming hardly portable without penalties.

We are proposing a new package intended as a toolkit to provide geometrical description of an experiment, full navigation functionality and additional tools to help building, checking and debugging geometries. Its development is a common effort of both ALICE off-line and ROOT teams that started 1.5 years ago.

Initially driven by ALICE needs related to the simulation and reconstruction framework, the new geometry is now designed as an experiment-independent package. Based on a GEANT-like architecture, it is able to represent and optimise the geometry performance of several HEP experiments. The gain in speed for navigation tasks ranges from 20% to about 800% compared to GEANT3. Since a significant fraction of the total simulation time is spent only for navigation purposes, this provides a valuable gain in performance.

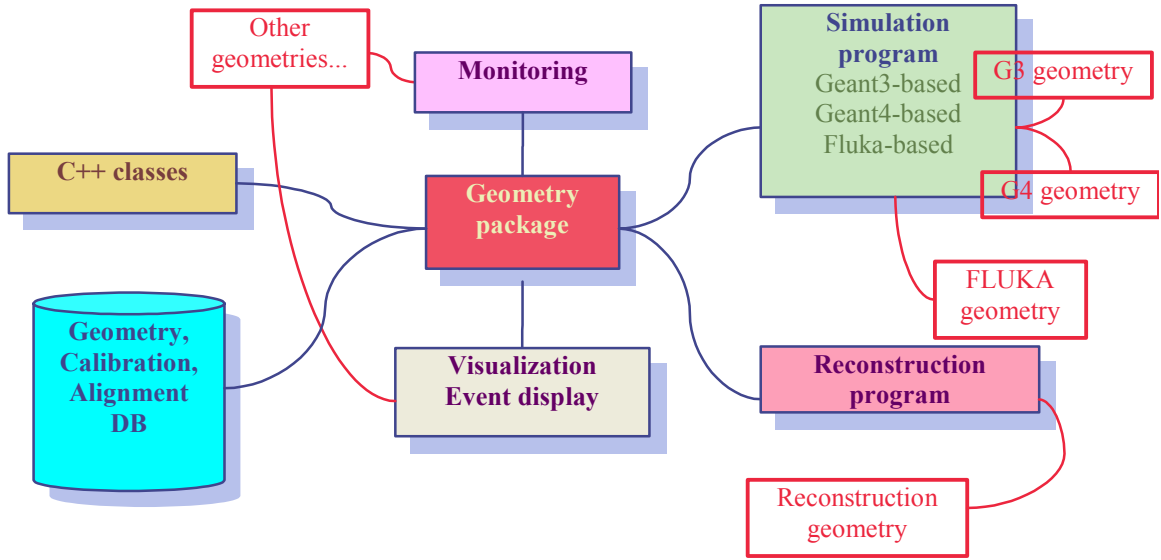

Figure 2: A single geometry available for all applications

The geometrical modeller is currently being integrated in a Virtual Monte-Carlo schema enabling running transparently several simulation MC's using the same user code.

## 2. ARCHITECTURE AND NAVIGATION

### 2.1. Design overview

The development of a new geometry package has started from a set of requirements and desired features:

- Provide basic navigation features like: *"Where am I in the geometry?" "How far from next boundary?"* or *"Which is the maximum safe step in any direction that does not cross any boundary?" "What is the normal to the boundary at the crossing point?".*
- Be able to map existing GEANT3 geometry descriptions in order to have a smooth transition from what already exist.
- Provide a compact implementation that scales with the number of objects in the geometry – both for memory management and performance.
- Improve tracking performance with respect to existing modellers.
- Provide full interactivity enabling users to easily build, debug and access relevant geometrical information.

Based on these requirements, we have adopted a GEANT-like architecture based on the container-contained concept [4] that was proven to work very well in case of detector geometries. The modeller (TGeo) [1] is developed within the ROOT framework [3] and currently provides a set of 16 basic shapes (primitives). It also provides a way of defining composite shapes as a result of Boolean operations between several primitives. Since the composition operation can be applied also to other composite shapes, this feature provides a quite large number of combinations covering most use cases. In order to facilitate geometry definition, the modeller provides also volume parameterisation methods as well as support for divisions to take advantage of detector symmetries.

The geometry is built by positioning volumes (a shape associated with a tracking medium) inside containers, which are as well volumes. Since volumes can be replicated several times, the resulting structure is a graph representing the logical geometric hierarchy. The only mandatory conditions in order to have the modelling features working properly are that positioned volumes (called nodes) should not extrude their containers and sister volumes inside the same container should not overlap each other. Since intended overlaps are sometimes unavoidable, users are allowed to declare them in order to be properly handled by the modeller.

Navigation inside geometries is optimised using volumetric divisions (voxels) at the level of the volumes. These structures allow minimising the number of node candidates that have to be checked for a given point and track direction, at the cost of additional memory.

### 2.2. Navigation features

A geometrical modeller has to provide answers to few basic queries related to geometry. Particle tracking is generally performed based on step management. At the heart of all simulation packages there is a physics engine able to generate the next process that will affect the currently transported particle and to propose the next step accordingly. Since the cross sections of most physics processes are highly dependent on material properties, the step manager has to know the current material, the distance to next boundary along current direction and the maximum safe step that can be made in the current volume. When crossing boundaries, some physics processes may also require the normal vector to the crossed surface.

This information can be retrieved from the API of the geometry manager class [4], which is free from any dependency on physics. We are currently developing specific interfaces for GEANT3, GEANT4 and FLUKA in order to be able to handle all geometry-related calls with the new geometry.

Most of the described navigation features are fully implemented and we have a working interface to GEANT3 package. Extensive testing against GEANT3 was done in order to validate navigation algorithms. For doing that, we have developed an automatic conversion tool from G3 to TGeo format. In this way we were able to port a large set of existing G3 geometries to the new modeller.

## 2.3. Validation procedure

Several geometries of existing HEP experiments were converted directly from GEANT3 ZEBRA banks. This insured a one to one mapping between G3 and ROOT geometry descriptions. We have performed GEANT3 simulations with the default physics setup of each experiment, collecting samples of one million points in the step manager. For each point we have saved additional information, like: current direction, current step length and safety distance, as computed by G3.

This information was saved both in HBOOK and ROOT formats and the point samples were than fed back into the two modellers. For each point we computed the predictions of both G3 and ROOT modellers. Proceeding this way, we were able to perform extensive test and debugging of the navigation algorithms. This led to several optimisations with respect to the early implementations.

Figure 3 shows an example of XY plot for all collected points. A consistency check was done as a first step: the predicted paths to the deepest node containing a given point were compared between GEANT3 and TGeo. This test typically gave differences ranging between few percents and 20-30% of the total number of points for high-complexity geometries. The reason for this comes from the fact that during a simulation, steps are always forced when the current particle crosses a boundary. Since the boundary positions of a given geometrical object are computed as a result of local matrix multiplication in the corresponding branch of the geometrical tree, these are always affected by a floating-point "diffusion" error increasing with the geometry granularity, e.g. the average number of levels per branch. Due to this reason, TGeo may classify points collected at boundaries on the other side than G3.

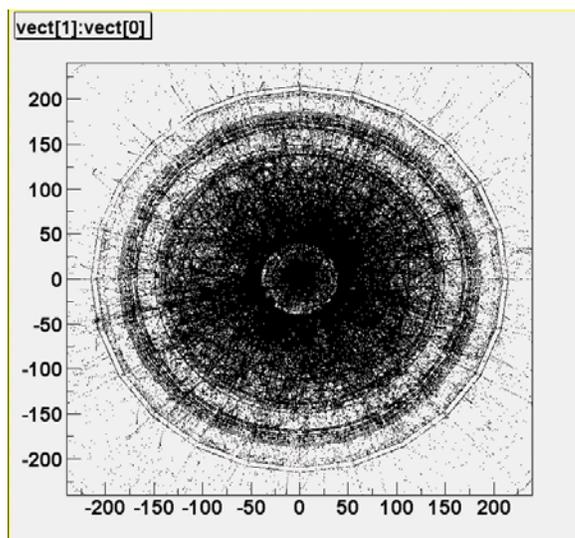

Figure 3: Points collected from G3 step manager.

We have checked this hypothesis by introducing a small smearing into the collected physics sample. This second sample was called "random" and typically reduced the "boundary effect" to less than 1% for all tested geometries. The few remaining non boundary-related differences were mostly coming from points belonging to the overlapping region of two or more positioned volumes. These can be equally classified as belonging to one or the other since it does not affect the physics results of a simulation when correctly defined. On the other hand, there were few cases when the answer of TGeo was different from G3 due to errors in the geometry definition (overlaps or extrusions). Since we have performed a point-by-point check of all differences, we were finally able to understand them and provide a detailed classification accordingly.

Similar validation tests were performed related to the computation of distances to the next boundary. We were also able to simulate events in the case of ALICE experiment having GEANT3 as main tracker and with TGeo performing computation of distances in parallel.

Finally we have made performance comparisons for the two modellers both for the physics and random samples. These tests show the average timing per point for different modellers in the case of several geometry setups. These represent either some simple GEANT3 examples or actual geometries of HEP experiments converted directly from G3.

The benchmarks are presented in Table 1 and show that TGeo modeller provides a large gain with respect to G3 in the cases where the later is unable to optimise its internal search algorithms; this usually happens for quite "flat" geometries where more than 500 volumes are positioned in the same container.

Table 1: TGeo performance vs. GEANT3 for *"Where am I?"* task. Point samples were collected during a G3-based simulation for each setup; random points come from smearing these samples. Averaged timings per point (in μs) and ROOT/G3 ratio are presented.

| Geometry | Nobjects | <nodes /volume> | FN_phys. G3 | FN_phys. ROOT | G3/ROOT | FN_rand. G3 | FN_rand. ROOT | G3/ROOT |
|---|---|---|---|---|---|---|---|---|
| Gexam1 | 425 | 0.17 | 1.18 | 0.73 | 1.62 | 2.41 | 1.18 | 2.04 |
| Gexam3 | 86 | 2.04 | 1.10 | 0.65 | 1.69 | 1.19 | 0.66 | 1.80 |
| Gexam4 | 12781 | 1.67 | 0.96 | 0.62 | 1.55 | 4.34 | 2.58 | 1.68 |
| ATLAS | 29046966 | 7.48 | 3.24 | 2.49 | 1.30 | 13.25 | 6.19 | 2.14 |
| CMS | 1166310 | 6.18 | 12.95 | 2.15 | 6.02 | 10.53 | 5.12 | 2.06 |
| BRAHMS | 2649 | 6.04 | 6.04 | 0.79 | 7.65 | 7.95 | 0.53 | 15.00 |
| CDF | 28525 | 11.45 | 25.04 | 1.90 | 13.18 | 5.60 | 1.07 | 5.23 |
| MINOS_NEAR | 30988 | 4.32 | 10.87 | 5.08 | 2.14 | 7.64 | 3.00 | 2.55 |
| BTEV | 295310 | 6.88 | 12.31 | 1.53 | 8.05 | 50.31 | 2.04 | 24.66 |
| TESLA | 15370 | 1.04 | 2.35 | 0.81 | 2.90 | 4.53 | 1.84 | 2.46 |

Performance for "Where am I" - physics case

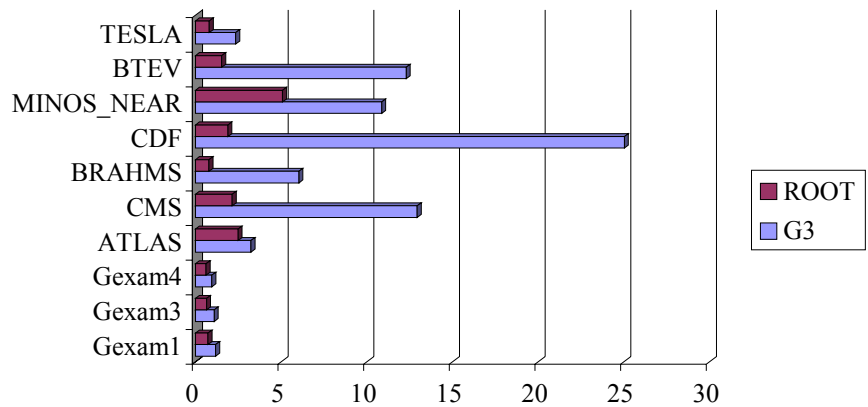

Performance for "Where am I" - random case

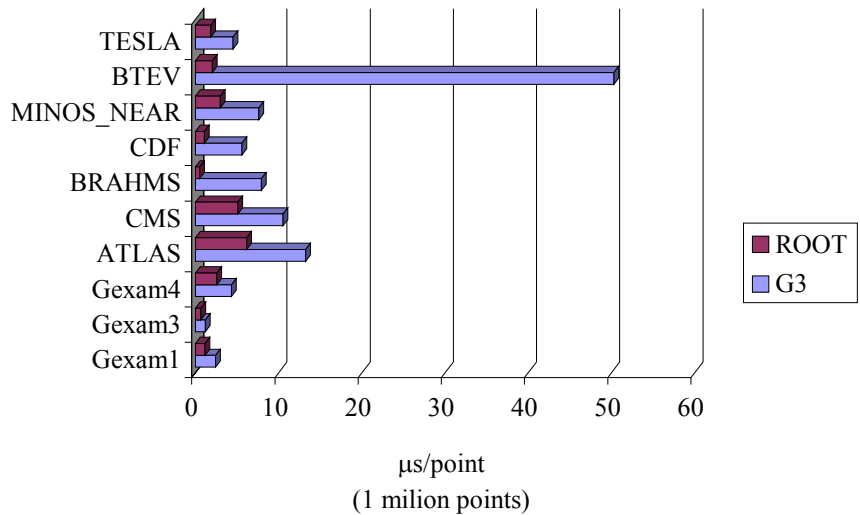

μs/point
(1 milion points)

Figure 4: Benchmarks for finding the deepest node in the geometry containing a given point ("Where am I?").

## 3. ADDITIONAL FEATURES

### 3.1. User Interface

The modeller user interface is implemented at two levels: all methods for building geometry, navigation interface utilities, visualization global settings and geometry checking are provided by the manager class [4]. Since any volume can become a top-level node in the geometry tree, there are several control methods and utilities (ray tracing, visualization, sampling) that are implemented at the level of volume objects.

Once geometry is successfully built and closed, there are several ways users can interact with it at graphics level. The geometry manager class is then registered and accessible in the ROOT browser [2]. The logical hierarchy of volumes can be fully parsed and all interface methods implemented at this level becomes accessible via context menus. The geometry-drawing package is part of a different library that is loaded on demand only when graphics-related calls are issued.

### 3.2. Visualization and checking utilities

Geometry visualization is currently implemented as a separate primitive-based interactive package allowing volume picking in parallel or perspective views. Hidden line/surface algorithms are supported for the time being by a separate package called X3D.

As the modeller architecture is based on embedding volumes into several layers, visualization supports a number of global options such as:

- Visualization of all volumes down to a given level;
- Visualization of final leaves in geometry, e.g. volumes having no daughters inside;
- Visualization of a specified branch or of a given volume only.

There are several visualization attributes that can be set directly at volume level:

- Visibility of the volume itself;
- Visibility of daughters;
- Colour and line attributes;

Once a picture is displayed, several methods can be used in order to perform validation checks for a given geometry module/branch:

- Sampling random points in the bounding box of the displayed volume branch. This method allows a direct visualization check of the modeller response for a given part of the geometry. Points classified by the modeller as belonging to volumes which are currently displayed are plotted with the same colour as the volume containing it;
- Isotropic ray tracing starting from a given point defined in the local reference frame of the displayed volume. Rays are fully tracked until they exit the current branch and only the segments crossing visible parts are displayed.
- Weight estimation of a given detector module. This is provided by a material sampling algorithm where it is possible to specify the desired precision.

In order to provide better interactivity, picked objects can be inspected. The viewing system provides navigation controls for zooming, moving, rotating and animating geometries.

### 3.3. Geometry checker

A simple but efficient geometry-checking tool was developed in order to be able to debug geometries. In fact, this tool is able to detect the vast majority of incorrect geometric constructs in reasonable time. As an example, a geometry like ALICE containing about 1.2 million positioned objects is fully checked within 30 seconds.

The checking algorithm verifies (within a given tolerance) if:

- Any of the positioned volumes extrude their container;
- Any of the volumes positioned inside the same container overlap each other, unless declared these volumes are declared overlapping.

The method loops over all pairs of candidate nodes, producing a list of illegal overlaps/extrusions that can be inspected with the ROOT browser. This list is sorted by decreasing overlap distance and all objects inside can be inspected and visualized on-line.

Table 2: Overlaps in ALICE geometry.

| Geometry | Overlaps/Extrusions | | |
|---|---|---|---|
| | > 1 mm | >100 μ | >10 μ |
| ALICE | 154 | 764 | 1460 |

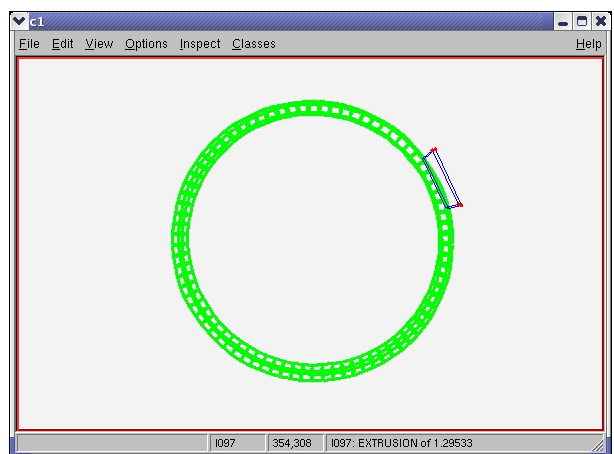

Figure 5: A typical extrusion

### 3.4. Geometry I/O

Currently the geometry modeller supports full ROOT I/O, meaning that a given geometry can be exported at any time as a ROOT file or retrieved back in a ready state. The size of the geometry file and the time to load are very important when designing and testing it.

In table 3 are presented the sizes of few geometry files as represented by TGeo modeller compared to the original size of the GEANT3 representation in RZ format. In order to have a more complete picture of the I/O performance, the time to read into memory each geometry file is presented against the total number of nodes.

Table 3: I/O load time and file size compared to GEANT3

| Geom. | No. Nodes | G3 .rz [Kb] | .root file [Kb] | CPU time PIII/800[s] |
|---|---|---|---|---|
| AMS | 112777 | 7372 | 4059 | 10.59 |
| ATLAS | 29046966 | 9863 | 4231 | 6.38 |
| BTEV | 295310 | 2048 | 839 | 1.30 |
| CDF | 24422 | 1818 | 1113 | 1.00 |

As a future plan, we consider the possibility to interface the geometrical modeller I/O with at least one relational database such as MySQL.

## 4. VIRTUAL MONTE-CARLO INTEGRATION

The ROOT geometrical modeller is currently being integrated in a Virtual Monte Carlo (VMC) schema [8] developed within the ALICE Software Project. The final goal is to be able to run transparently several simulation engines using the same geometry. Being independent from any simulation package, this geometry can also be used in a reconstruction code.

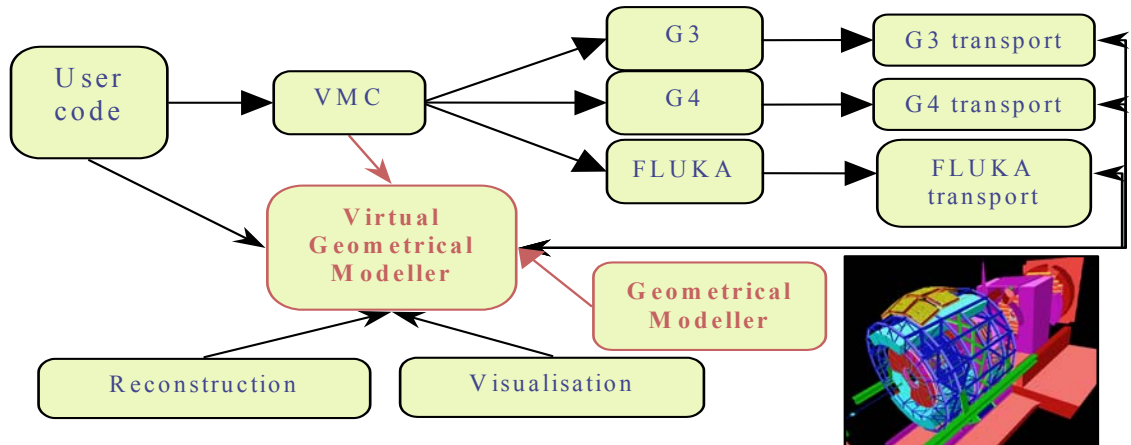

Figure 6: Schematic view of the experimental software framework based on the VMC and an external geometrical modeller

The VMC is a set of abstract interfaces that provides access for user code to the data structures (common blocks in case of FORTRAN or classes for C++) and methods of the simulation programs behind it. It allows user Monte Carlo applications to access in a common way GEANT3, GEANT4 or FLUKA.

The pure virtual methods in the VMC are implemented by specific interfaces to the simulation programs. Among several methods for controlling simulation behaviour, step management and particle stack, there are specific methods for building and retrieving geometrical information. In the current production version of VMC, these methods are implemented inside the simulation specific interfaces and they build the native geometries specific to the simulation packages used.

In order to provide a single geometry to be used by all simulation programs, we need

- To provide the implementation of the VMC methods for building and accessing TGeo geometry which can then replace the current implementation for building native geometries.
- To replace the navigation functionality of native geometries with corresponding features provided by

TGeo by redirecting all navigation queries to the new geometry.

This is possible in case of GEANT3 and FLUKA by wrapping modelling specific subroutines. In case of GEANT4 [7] this would be also eventually possible if the package will provide an abstract layer for navigation. This layer should contain the prototypes for all basic queries like: master-to-local transformations, computation of distances to boundaries, safety distance or computation of normal vectors to crossed surfaces. GEANT4 step manager as well as all other G4 objects that are related to navigation would have to rely only upon calls to this abstract layer. In this case the actual G4Navigator class will implement the navigation functionality for native G4 geometry descriptions, while in case of TGeo we will need to provide a specific implementation. Going the other way around, a version of a TGeo to G4 geometry converter is under development, so that G4 users will be able to run GEANT4-based simulations starting from a ROOT geometry description [8].

Currently the version of GEANT3 with direct use of TGeo is working. GEANT3 tracking can be fully done without filling G3 geometry common blocks but only by building its materials and media. The only remaining step for the full implementation of a production version of VMC based on TGeo package is the computation of the normal vectors to the crossed surfaced, required by optical physics processes. This feature is currently under development.

FLUKA [6] is a simulation program using different concepts than GEANT related to the geometry description. Since the VMC design was inspired by GEANT3, realistic detector geometries would be very difficult to build by using directly FLUKA geometry package. However, an interface between FLUKA and GEANT4 geometry can be currently used to run FLUKA driven by G4 navigation. We are currently in the process of designing a similar one based on TGeo modeller.

An important aspect of the version of the VMC working with the new package is that users will be able to build the geometry by using directly the API of TGeo. The geometry built in this way can be stored in ROOT format and then directly retrieved and used it the context of VMC.

## 5. CONCLUSIONS

A new geometrical modeller able to represent a large number of HEP experiments is being developed by ALICE and ROOT teams.

This will provide a unique representation of ALICE geometry and will allow running transparently several MC's and reconstruction programs using not only the same geometry description, but also the same geometry modeller.

Performance was the highest priority during the development and this is reflected by the benchmarks. The code is available in the ROOT distribution.